\documentclass[prl, twocolumn, superscriptaddress, reprint, aps, floatfix]{revtex4-1}

\pdfoutput=1

\usepackage{empheq}
\usepackage{amsmath} 
\usepackage{braket}
\usepackage{amsfonts}
\usepackage{graphicx} 
\usepackage{float}
\usepackage{bm}
\usepackage{color}
\definecolor{bleuf}{rgb}{0,0.44,0.72}
\usepackage[unicode=true,pdfusetitle,bookmarks=true,
bookmarksnumbered=false,bookmarksopen=false,breaklinks=false,
pdfborder={0 0 0},backref=false,colorlinks=true,citecolor=bleuf,
linkcolor=bleuf,urlcolor=bleuf]{hyperref}

\usepackage{cleveref}
\usepackage{soul}

\newcommand{\toggl}[1]{} 

\begin{document}

\title{Motile dislocations knead odd crystals into whorls }  

\author{Ephraim S. Bililign}
\affiliation{James Franck Institute  and Department of Physics, University of Chicago, Chicago, IL 60637, USA}

\author{Florencio Balboa Usabiaga}
\affiliation{Basque Center for Applied Mathematics (BCAM), Alameda de Mazarredo 14, 48009 Bilbao, Spain}
\affiliation{Center for Computational Biology, Flatiron Institute, New York, NY
10010} 

\author{Yehuda A. Ganan}
\affiliation{James Franck Institute  and Department of Physics, University of Chicago, Chicago, IL 60637, USA}

\author{Alexis Poncet}
\affiliation{University of Lyon, ENS de Lyon, University of Claude Bernard, CNRS, Laboratoire de Physique, F-69342 Lyon, France}

\author{Vishal Soni}
\affiliation{James Franck Institute  and Department of Physics, University of Chicago, Chicago, IL 60637, USA}

\author{Sofia Magkiriadou}
\affiliation{James Franck Institute  and Department of Physics, University of Chicago, Chicago, IL 60637, USA}

\author{Michael J. Shelley}
\email{shelley@cims.nyu.edu}
\affiliation{Center for Computational Biology, Flatiron Institute, New York, NY
10010} 
\affiliation{Courant Institute, NYU, New York, NY 10012}

\author{Denis Bartolo}
\email{denis.bartolo@ens-lyon.fr}
\affiliation{University of Lyon, ENS de Lyon, University of Claude Bernard, CNRS, Laboratoire de Physique, F-69342 Lyon, France}

\author{William T. M. Irvine}
\email{wtmirvine@uchicago.edu}
\affiliation{James Franck Institute, Enrico Fermi Institute and Department of Physics, University of Chicago, Chicago, IL 60637, USA}

\maketitle

{\bf The competition between thermal fluctuations and potential forces governs the stability of matter in equilibrium, in particular the proliferation and annihilation of topological defects. 
However, driving matter out of equilibrium allows for a new class of forces which are neither attractive nor repulsive, but rather transverse. 
The possibility of activating transverse forces raises the question of how they affect basic principles of material self-organization and control. 
Here, we show that transverse forces organize colloidal spinners into odd elastic crystals crisscrossed by motile dislocations. 
These motile topological defects organize into a poly-crystal made of grains with tunable length scale and rotation rate. 
The self-kneading dynamics drive super-diffusive mass transport, which can be controlled over orders of magnitude by varying the spinning rate. 
Simulations of both a minimal model and fully resolved hydrodynamics establish the generic nature of this crystal-whorl state. 
Using a continuum theory, we show that both odd and Hall stresses can destabilize odd-elastic crystals, giving rise to a generic state of crystalline active matter. Adding rotations to a material’s constituents has far-reaching consequences for continuous control of structures and transport at all scales.
} 

The celebrated interplay between configurational entropy and the energetics of topological defects in two-dimensional melting have provided a lens through which to understand the phases of  condensed matter~\cite{chaikin}, such as superfluid films,  colloids and liquid crystals \cite{Strandburg,Bishop1978,Zahn1999,alsayed2005premelting, meng2014elastic,Thorneywork2017}.
While these  systems span a wide range of particle interactions, scales, and intermediate phases, they are all unified in that the forces between constituents are primarily longitudinal, their dynamics are equilibrium, and their interactions are symmetric under both time-reversal and parity. 
What happens if the inter-particle interactions include transverse forces as well ~\cite{Cafiero2002,grzybowski_dynamic_2000,yan_rotating_2014,nguyen_emergent_2014,yeo_collective_2015,kokot_emergence_2015,soni2019odd,Shen2020,Liu2020}? 
This deceivingly minimal generalization can break these assumptions at a fundamental level.

In equilibrium,  transverse forces cannot alter the phase behavior of condensed matter.
However, there is no such guarantee out of equilibrium and such transverse interactions generically occur in collections of naturally and artificially spinning objects. 
Examples include planetary disks \cite{armitage1998turbulence}, spinning cell aggregates and
membrane inclusions \cite{petroff2015fast,OSS2019}, active colloids and grains \cite{nguyen_emergent_2014,van_zuiden_spatiotemporal_2016,aubret2018targeted,snezhko_complex_2016, scholz_rotating_2018, kokot2017,Shen2020,Liu2020,lim2019cluster}, atmospheric scale dynamics \cite{bouchet2012statistical, baroud2002anomalous}, parity-breaking  fluids \cite{korving1966transverse, hoyos2014effective, wiegmann_anomalous_2014, souslov2019topological, banerjee_odd_2017, han2020statistical} and simple models of turbulence \cite{eyink2006onsager}.
How does this more general and ubiquitous form of matter generically self-organize, what are its stable phases, and how does it transition between them?

\begin{figure*}
\centering
\includegraphics[width=\textwidth]{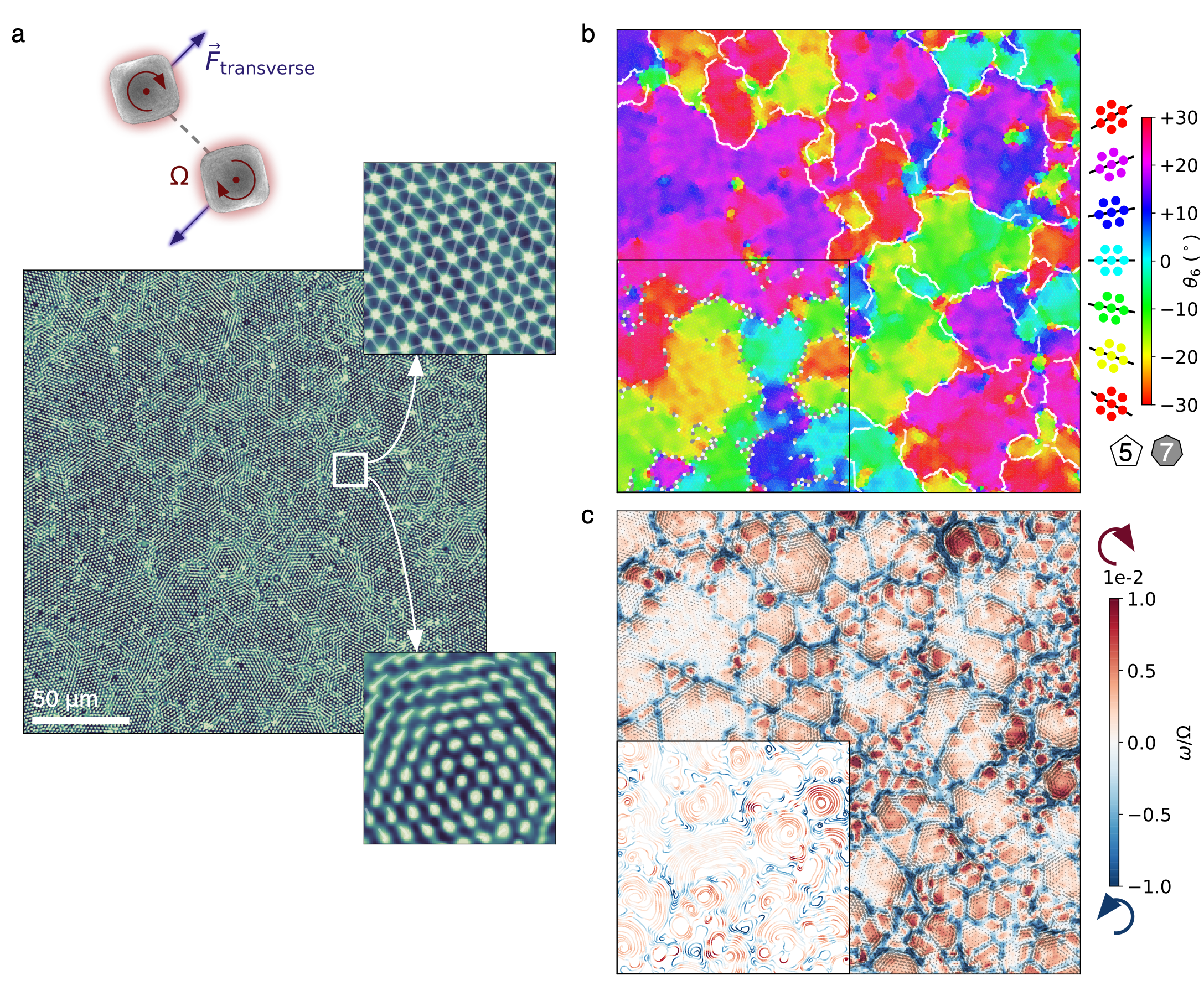}
\caption{\label{fig:polycrystal} {\bf A crystal whorl state.} 
{\bf a,} A dense and dynamic phase of colloids spinning at a frequency $\Omega$ and interacting through both longitudinal and transverse pairwise interactions is directly imaged with a microscope through crossed polarizers. The rotation-averaged position of each particle appears as a bright spot and reveals intermittent crystalline order (see Supplementary Movie 1). Magnifying a region reveals a highly-ordered crystalline structure (top), and time-averaging further reveals a rotating flow (bottom).
{\bf b,} To further illuminate the polycrystalline structure of this phase, we color particles by the angle of the local bond-orientational order parameter $\theta_6$. 
The polycrystal can be segmented into domains, and boundaries are drawn between them
(see Supplementary Movie 2).  
The inset highlights the individual defects between grains, underlying grain boundaries.
{\bf c,}  The polycrystal is dynamical and displays intermittent  vortical flows as revealed when each region is colored by its vorticity $\omega$ (see Supplementary Movie 3). The inset presents the same vortical information through the streamlines of the particle flow (see Supplementary Movie 12).}
\end{figure*}

Figure~\ref{fig:polycrystal}a and Supplementary Movie 1
show a ${\sim}200\times 200\,\mu \rm m$ region within a centimetre-scale monolayer of magnetic colloids.
Each particle is uniformly spun by an externally applied magnetic field, resulting in their self-organization into a dynamic and dense phase.
The active rotation of the magnets gives rise to both longitudinal magnetic attraction and sustained chiral transverse hydrodynamic interactions, as illustrated in the inset of Figure~\ref{fig:polycrystal}a.
Crucially, the forces are separation-dependent and can be tuned by varying the rotation frequency, providing an ideal platform for exploring how transverse interactions shape the dense phases of chiral matter.

\begin{figure*}
\centering
\includegraphics[width=\linewidth]{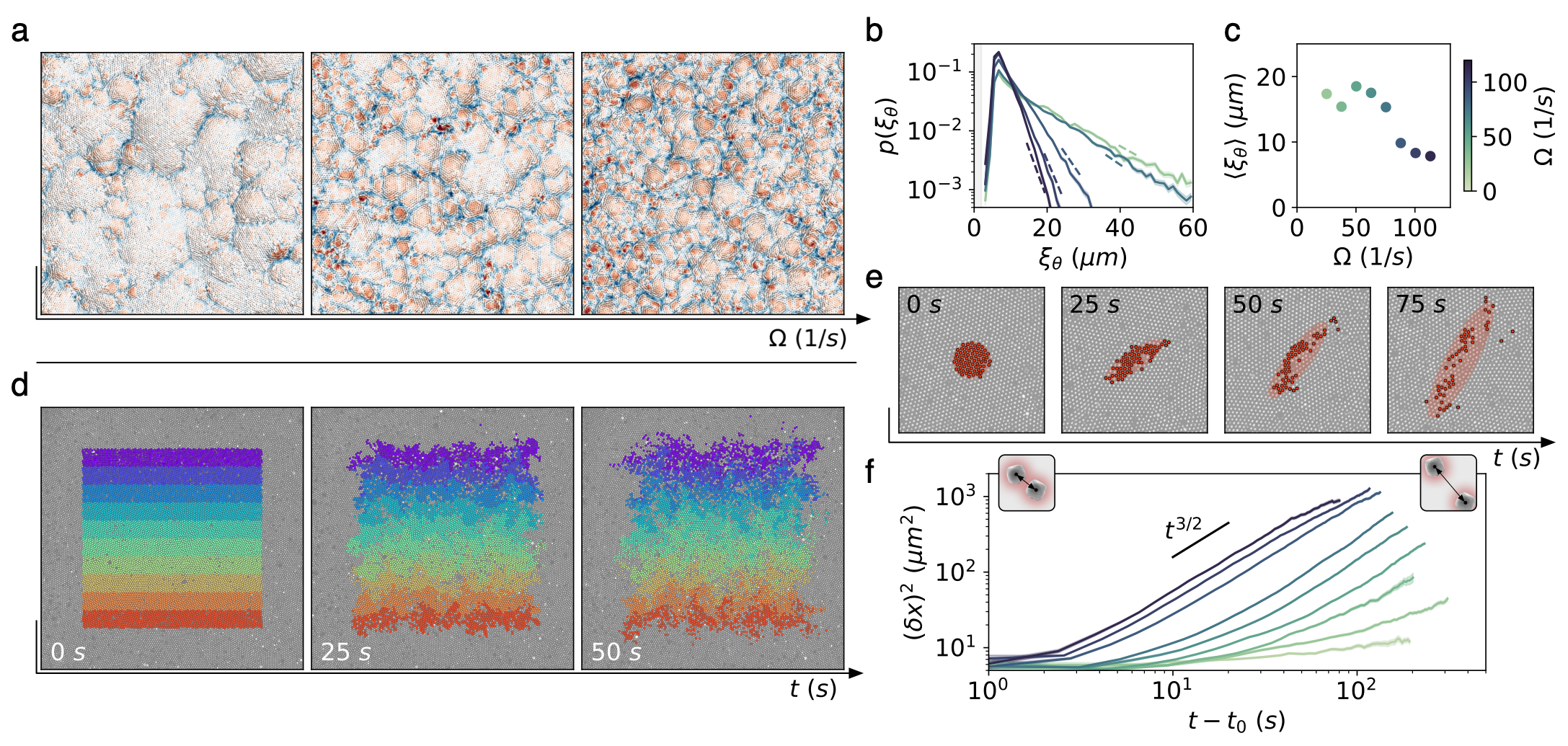}
\caption{\label{fig:odd} {\bf Transport in the crystalline whorl state.}
{\bf a,} The self-kneading of crystalline patches of material is tunable through the rotation frequency to yield a range of scales. Same colormap as in Figure~\ref{fig:polycrystal}c. {\bf b,} These states are characterized by an exponential distribution of grain sizes, dashed straight lines represent the slope of the associated exponential fit.
{\bf c,} Accordingly, the average size of grains in the steady-state tends to decrease with frequency. Error bars, which are smaller than markers, represent standard error.
{\bf d,} The constant structural kneading of the chiral whorl state by topological defects introduces novel mixing properties which can be imaged by artificially dying stratified layers in a crystal that subsequently bleed into each other.
{\bf e,} By contrast to conventional diffusive processes, the smearing of the fluid over time is a strongly anisotropic process, wherein an example blob of fluid is pulled apart by the flow between two chiral whorls.
{\bf f,} The pairwise separation $(\delta x)^2$ plotted versus time $t$ for particles initially in close proximity suggests that this abnormal spreading gives rise to superdiffusive behavior, which itself is a function of rotational frequency. Error bars represent standard error.
}
\end{figure*}

Upon spinning our particles, we find that the system generically self-organizes into crystal `whorls'.
A snapshot, colored by the phase of the crystalline bond-orientational order parameter, $\psi_6(\bm{x})$ (see Figure~\ref{fig:polycrystal}b and Supplementary Movie 2), reveals a polycrystalline arrangement of grains of triangular crystal order separated by topological defects organized into grain boundaries~\cite{lavergne2017anomalous}.
This picture is reminiscent of metallurgical crystalline phases with quenched disorder; however, unlike their crystalline static counterpart, the structure is continually evolving, grain boundaries move, collapse, and spontaneously emerge as crystalline domains rotate like vortical whorls (see Figure~\ref{fig:polycrystal}c and Supplementary Movie 3, 12).
Segmenting the phase into crystalline domains enables us to study its statistical properties, revealing that after a short transient, the domains within this polycrystal settle to a constant characteristic size (SI Sec.~2.2).
This scale can be tuned by altering the particle rotation rate alone, yielding either significantly larger or smaller crystalline whorls, as in Figure~\ref{fig:odd}. 

\begin{figure*}
\centering
\includegraphics[width=\textwidth]{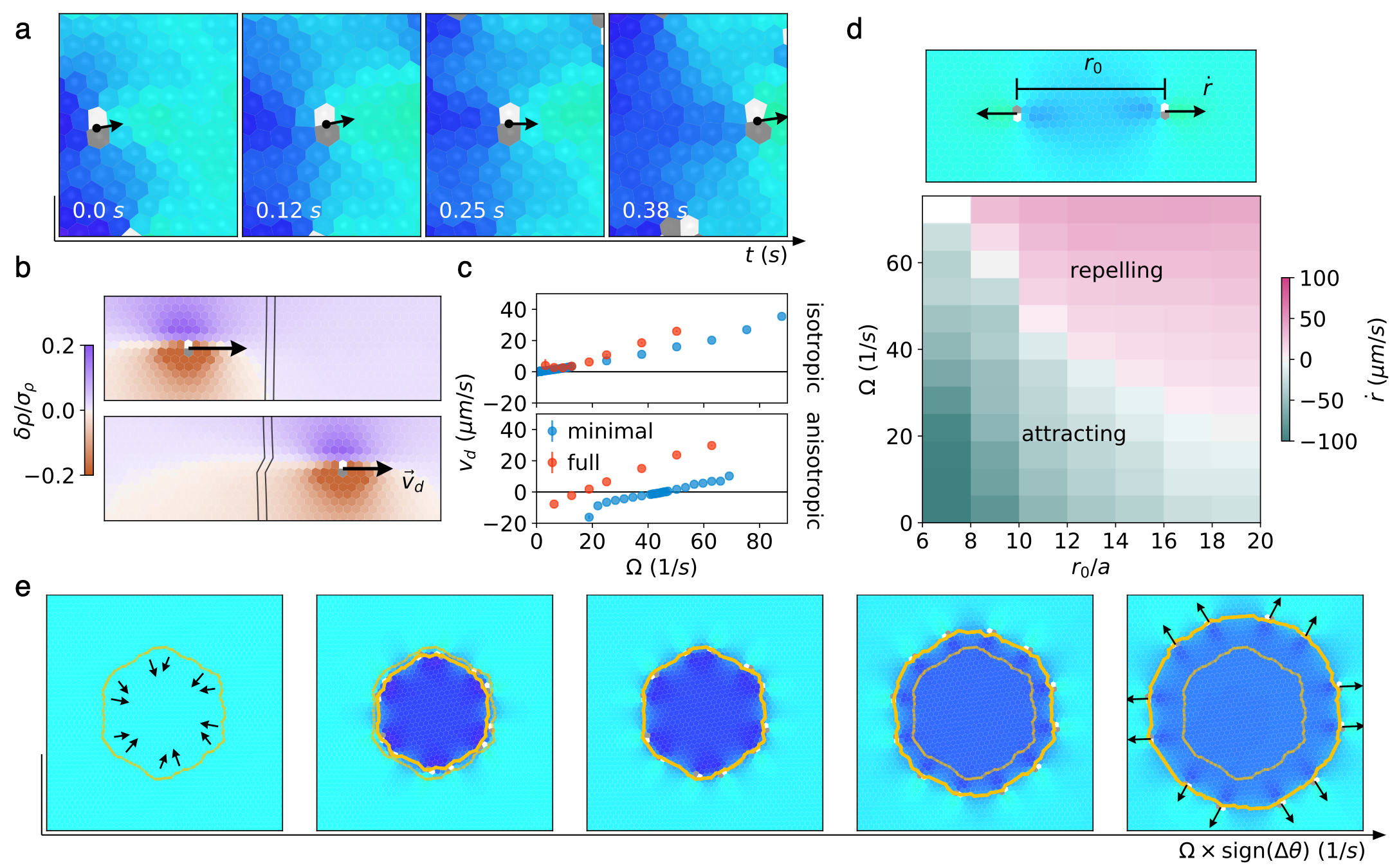}
\caption{\label{fig:transport}{\bf Motile dislocations.} 
Dislocations in the chiral crystalline phase behave like active particles.
 {\bf a,} In the experiment, dislocations are observed to move ballistically in the direction of their Burgers vector. Same colormap as in Figure~\ref{fig:polycrystal}b.
{\bf b,} This behavior is reproduced in simulations of both the full hydrodynamic  and minimal models by initializing a configuration corresponding to a single dislocation in an otherwise undefected crystal.
In a crystal of particles interacting via transverse forces, we can intuit the dislocation's direction of motion from the relative displacement of the crystal, which is colored by the relative density $\delta\rho = \rho-\bar{\rho}$ normalized by the standard deviation $\sigma_\rho$. 
{\bf c,} The precise dislocation speed depends weakly on the details of the interactions. In both the minimal and full hydrodynamic models, the speed increases with frequency for isotropic dipole interactions (top). By contrast, dislocation motility is reversed at a frequency threshold for anisotropic dipole interactions (bottom). Error bars represent fit covariance (minimal) and standard error (full).
{\bf d,} By tuning the frequency and initial separation, two defects that would otherwise attract and annihilate can be made to repel, overwhelming even elastic forces with transverse ones. Same colormap as in Figure~\ref{fig:polycrystal}b.
{\bf e,} The collective dynamics of many defects arranged to form a grain boundary inherits this sensitivity to transverse forces. Such a grain boundary collapses when elastic forces dominate, and expands without bound when transverse forces dominate. Same colormap as in Figure~\ref{fig:polycrystal}b, orange line indicates position of grain boundary in both initial and present state for comparison.
}
\end{figure*}

What powers this lively steady state?
Careful inspection reveals that the motion of topological defects in the crystalline structure is unlike the familiar motion of dislocations found in conventional passive materials. 
Conventional dislocations are either stationary or diffuse bi-directionally driven by thermal fluctuations; we observe instead that  in our chiral medium they move ballistically as seen in Figure~\ref{fig:transport}a and Supplementary Movie 4.

We can gain an intuitive understanding of what powers dislocation motility   by inspecting  the plastic  deformations of the crystal brought about by dislocation glide. 
As shown in Figure~\ref{fig:transport}b,
dislocation glide reflects the displacement of one crystal plane over the other. 
As this deformation is equal parts rotation and shear, it is   naturally actuated by the  rotational drive.

To isolate this phenomenon, 
we introduce a minimal model of the overdamped dynamics of spinners interacting via transverse,~\nocite{ANDERSON2020109363} frequency-dependent forces and potential longitudinal forces (SI Sec.~4.1). 
This minimal approach is informed by full hydrodynamic simulations of particles that closely approximate our experimental system (SI Sec.~5.1).
The transverse forces arise primarily from near-field hydrodynamic interactions, while the longitudinal interactions arise  primarily from both  steric repulsion and  magnetic attraction (SI Sec.~5.5).

By initializing simulations with a single dislocation in an otherwise perfectly ordered triangular crystal (see Supplementary Movies 5-6), we are able to observe the motion of dislocations isolated from interactions with other defects in both minimal and full hydrodynamic simulations.
As shown in Figure~\ref{fig:transport}c, we observe that the glide speed is frequency dependent. 
When the longitudinal interactions between particles are isotropic, corresponding to time-averaged magnetic interactions (SI Sec.~4.5, 5.3), the glide velocity is a monotonically increasing function, with a  threshold (Figure~\ref{fig:transport}c, SI Sec. 4.5.1). This is consistent with the notion of a unidirectional propulsion resisted by local Peierls barrier~\cite{peierls1939reports, nabarro1947dislocations}. 
In the case of anisotropic dipolar interactions, at low frequencies we observe a correction to defect propulsion brought about by a competition between magnetic and rotational interactions (Figure~\ref{fig:transport}c).

The motility of individual dislocations provides a significant twist on the collective dynamics of defects normally driven by elastic interactions~\cite{Schall2004,weinberger2008surface, Irvine15544,amir2012dislocation,Deutschlander2015,Braverman2020}. 
For example two defects that would normally attract and annihilate in response to elastic forces, can instead unbind when propelled by transverse forces as shown in Figure~\ref{fig:transport}d.

The collective dynamics of several defects is similarly affected. 
Figure~\ref{fig:transport}e shows snapshots from simulations in which we initialized a finite size grain in an otherwise perfect crystal and varied the rotation frequency (see Supplementary Movies 7-8). 
Altering the frequency affects transverse interactions most strongly and thus enables us to tune the balance between stabilizing elastic interactions and defect motility. 
As shown in Figure~\ref{fig:transport}e, 
the grain size is set by the competition between motility and defect interactions. 
At low rotation frequencies the grains are stable, and their size  becomes larger when the defect's propulsion direction is outwards. Similarly, they shrink and collapse for inwards defect motility. 
When the system is driven sufficiently strongly, the defect motility overpowers the elastic interactions resulting in unstable grain boundaries.

In the polycrystalline state we observe in our experiments, the instability of grain boundaries results in the exchange of dislocations between adjacent grains, as well as defect proliferation. 
These  nonlinear dynamics drive the system to the dynamical crystal whorl state erasing any memory of the initial configuration. 

The flow field $\bm{v}(\bm{x}_i)$ that emerges from the combination of motility and proliferation can be readily measured from individual particle trajectories and its corresponding vorticity $\omega(\bm{x}_i)$ is shown in Figure~\ref{fig:polycrystal}c. 
When time-averaged, this flow field  corresponds to a quiet bulk (SI Sec. 2.3) and a lively edge~\cite{soni2019odd}.
However our time-resolved measurements reveal instantaneous dynamics shaped by unsteady vortical flows  illustrated in Figure~\ref{fig:polycrystal}c. %
The dynamical and structural pictures of this chiral whorl state are aligned.
As seen in Figure~\ref{fig:polycrystal}b-c, the grain boundaries support strongly localized flows having a vorticity opposite to the particle rotation.
By contrast, the grains correspond to low positive vorticity, intermittently interrupted by isolated dislocations zipping through.

In this chaotic stationary state, the balance between even and odd forces remains the controlling parameter that determines the characteristic size of the chiral whorls, which can be viewed through the vorticity in Figure~\ref{fig:odd}a and the distribution of crystalline grains in Figure~\ref{fig:odd}b-c (SI Sec. 2.2-2.3, 4.4, 5.2).
The sustained proliferation and annihilation of motile dislocations is reminiscent of active nematics where motile disclinations power spatiotemporal chaos~\cite{Dogic2012}. 
Here dislocations give rise to self kneading crystal whorls.

This self-kneading of the crystal phase results in enhanced mixing which can be qualitatively captured  by artificially tagging the colloids and watching them spread; see Figure~\ref{fig:odd}d and Supplementary Movie 11. 
As shown in Figure~\ref{fig:odd}e, an artificially dyed blob spreads anisotropically before disintegrating into separate blobs, hinting at a mechanism reminiscent of Richardson diffusion in turbulence. 
We investigate this quantitatively by tracking the mean squared separation between pairs of particles in the chiral phase. Figure~\ref{fig:odd}f shows that pair separation is super-diffusive above a separation that corresponds to the characteristic grain size due to a punctuated mix of conventional diffusion within crystalline whorls and Richardson-like diffusion between them. 
Note that tuning the frequency of rotation alters both the basic unit of time as well as the domain size (Figure~\ref{fig:odd}d and e) enabling effective diffusion rates to be tuned over orders of magnitude.

\begin{figure*}
\centering
\includegraphics[width=\linewidth]{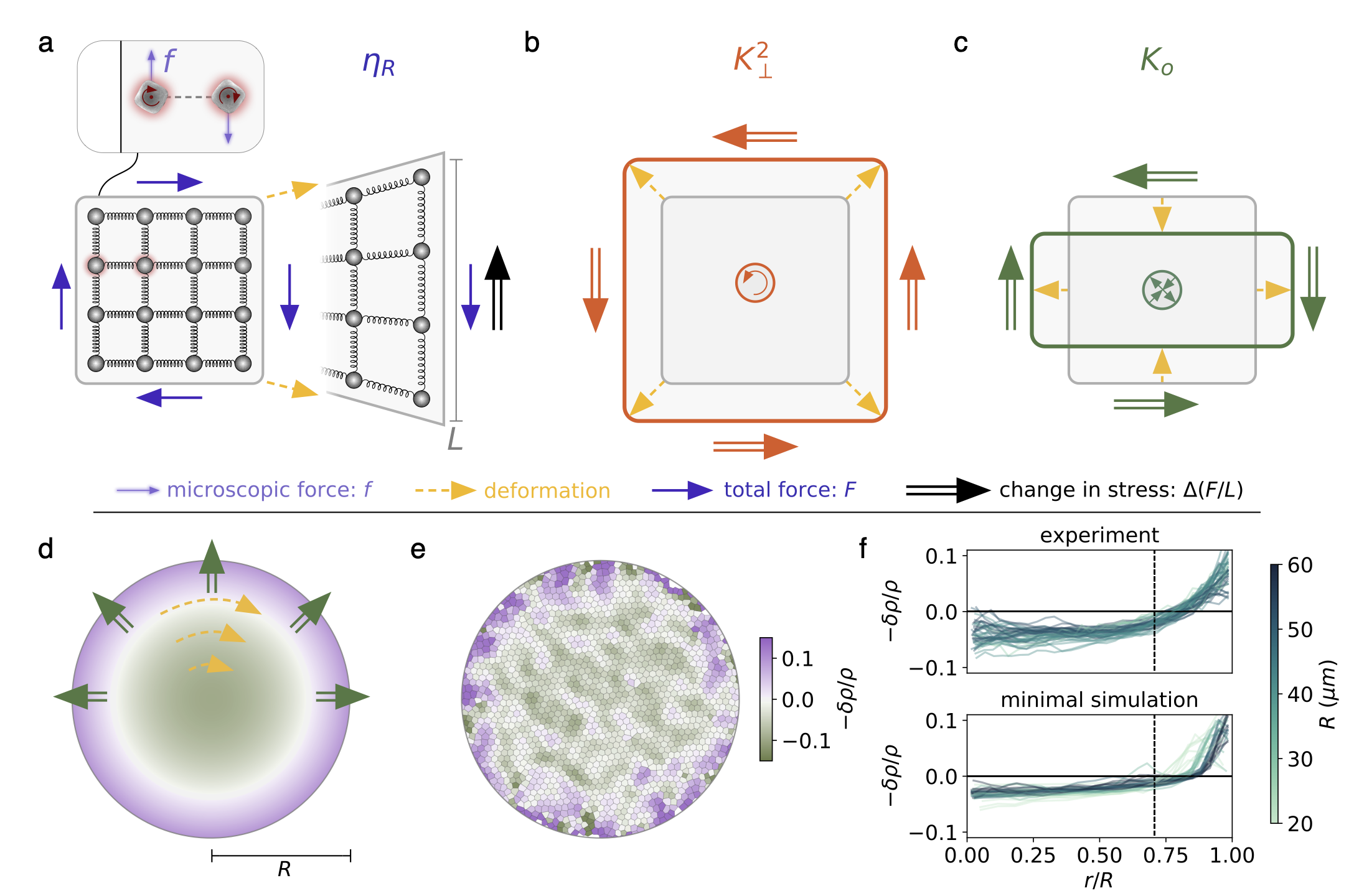}
\caption{\label{fig:moduli} {\bf Odd response in the steady-state.} 
{\bf a,} A patch of material built of  particles that interact with their neighbors  via constant, separation-independent transverse forces $f$ parallel to the material's edge sustains a net odd stress $2\eta_R\Omega=n f/L$ on its edge, where $n$ is the number of particles on an edge of length $L$ (see SI Sec. 6.2 for a more general discussion). On each edge, the macroscopic total force $F=nf$  is unchanged upon dilating the  edge, but as the length $L$ of the boundary is increased, there is a net change in stress $\Delta (F/L)$ in a direction opposite to the odd stress.
{\bf b,} For a uniform dilation, the deformation leads to a net counter-rotational stress. This coupling between dilation and rotation is denoted by $K_\perp^2$ in the elasticity tensor.
{\bf c,} Likewise, for a shear  deformation, the system acquires a net rotated shear stress, representative of the odd elasticity $K_o$.
{\bf d,} In a droplet of crystalline material, the odd stress induces a net rotation, balanced by friction with the substrate. 
The naturally occurring azimuthal distortion generates, via the odd elastic moduli, a characteristic radial density dependence (SI Sec.~7.2).
{\bf e,} We observe this signature of odd elastic response in experiments on droplets by measuring the dilational component of the strain through the relative density $\delta\rho/\rho$.
{\bf f,} In both the experiments and simulations, $\delta\rho/\rho$ transitions radially within a droplet from contracting to dilating in a manner independent of droplet radius $R$. In both cases, we obtain excellent qualitative agreement with the theoretical prediction for the transition radius, denoted by the dashed line (SI Sec.~7.2).
}
\end{figure*}

Is the crystal whorl state generic? 
We repeated the experiments shown in Figure~\ref{fig:polycrystal} while varying the applied magnetic field strength, adding a static vertical component of magnetic field, and varying the shape of the magnetic particles (SI Sec.~2.6). 
Additionally,  we varied the microscopic interactions in our minimal model simulations (SI Sec.~4.8).
Finally, in full hydrodynamic simulations, we simulated spheres and cubes that interact both magnetically and via isotropic attraction potentials (SI Sec.~5.2). 
In all cases  we find that when transverse interactions dominate,  the systems self-organizes into  a crystal whorl state, supporting the notion that this state is generic.
To gain further insight into the origins of this  new state of matter, we adopt a continuum perspective and investigate destabilization mechanisms within this approach. 
As illustrated in Figure~\ref{fig:moduli}, the microscopic rotational drive induces transverse interactions between the constituent particles which in turn give rise to active stresses in the crystal phase. 
To gain an essential insight, we first consider the simplified case of a material in which neighboring particles  interact via constant pairwise transverse forces. 
Figure~\ref{fig:transport}, shows how such interactions naturally give rise to a uniform anti-symmetric odd stress~\cite{dahler_theory_1963,tsai_chiral_2005,soni2019odd}, as well as all  elastic moduli  that can arise in an isotropic solid when energy conservation cannot be assumed~\cite{scheibner2020odd}.
As shown in Figure~\ref{fig:moduli}b-c, the moduli originate geometrically through the  changes in perimeter of an infinitesimal material patch. A dilation (b) does not alter the total force on the edge, but increases the perimeter, thereby coupling dilation and rotational stress. Similarly, a shear  deformation (c) alternately increases and decreases the length of the  edges giving rise  to a rotated shear stress (SI Sec.~6.2).

The addition of spatial dependence to the interaction forces (SI Sec.~6.3) makes estimating the moduli  more challenging, however it  does not alter the basic conclusion that a crystal of spinning particles is a quintessential odd  elastic solid.
Measuring the deformation field in a collection of nearly-circular, mono-crystalline droplets (Figure~\ref{fig:moduli}e-f) provides corroborating evidence from experiments and simulations. 
The droplets rotate at a constant angular velocity that decreases with their radius $R$. Regardless of the droplet size, they consistently display a radial strain profile that varies from  compressive in the center to dilational on the edge.
Within an elastic description, this qualitative feature must originate from odd-elastic moduli.
A continuum prediction of the strain field (SI Sec.~7.2) confirms this intuition and  predicts the shape of the density profile, as illustrated in Figure~\ref{fig:moduli}d.

A natural question  to ask is then whether odd elastic solids powered by odd stresses are linearly stable.
We consider the continuum description of an elastic solid with a local displacement $\bm{u}(\bm{r},t)$ and local velocity $\bm{v}(\bm{r},t)$, that is allowed to experience all stresses consistent with broken parity and time-reversal (SI Sec.~6.1, 8.1)~\cite{soni2019odd,scheibner2020odd}. 
In addition to elastic contributions, the symmetric stress, includes even and odd  viscous contributions $\sigma_{ij} = -p\delta_{ij} + K_{ijkl} \partial_k u_\ell+ \eta_{ijkl} \partial_k v_\ell$. 
The inner drive imposed by the spinners is encoded by the  antisymmetric stress $\sigma_{ij}^{\mathrm{spin}} = 2\eta_R \epsilon_{ij}\Omega$.  
Ignoring inertia, the dynamics is given by the balance between  viscoelastic stresses and substrate drag defined by a constant friction coefficient $\Gamma$:
\begin{equation}
\Gamma \bm{v} =  \nabla\cdot \bm{\sigma}+\nabla\cdot \bm{\sigma}^\mathrm{spin},
\end{equation}
and the continuity equation $\partial_t \rho + \nabla\cdot (\rho\bm{v}) = 0$.

\begin{figure*}
\centering
\includegraphics[width=\linewidth]{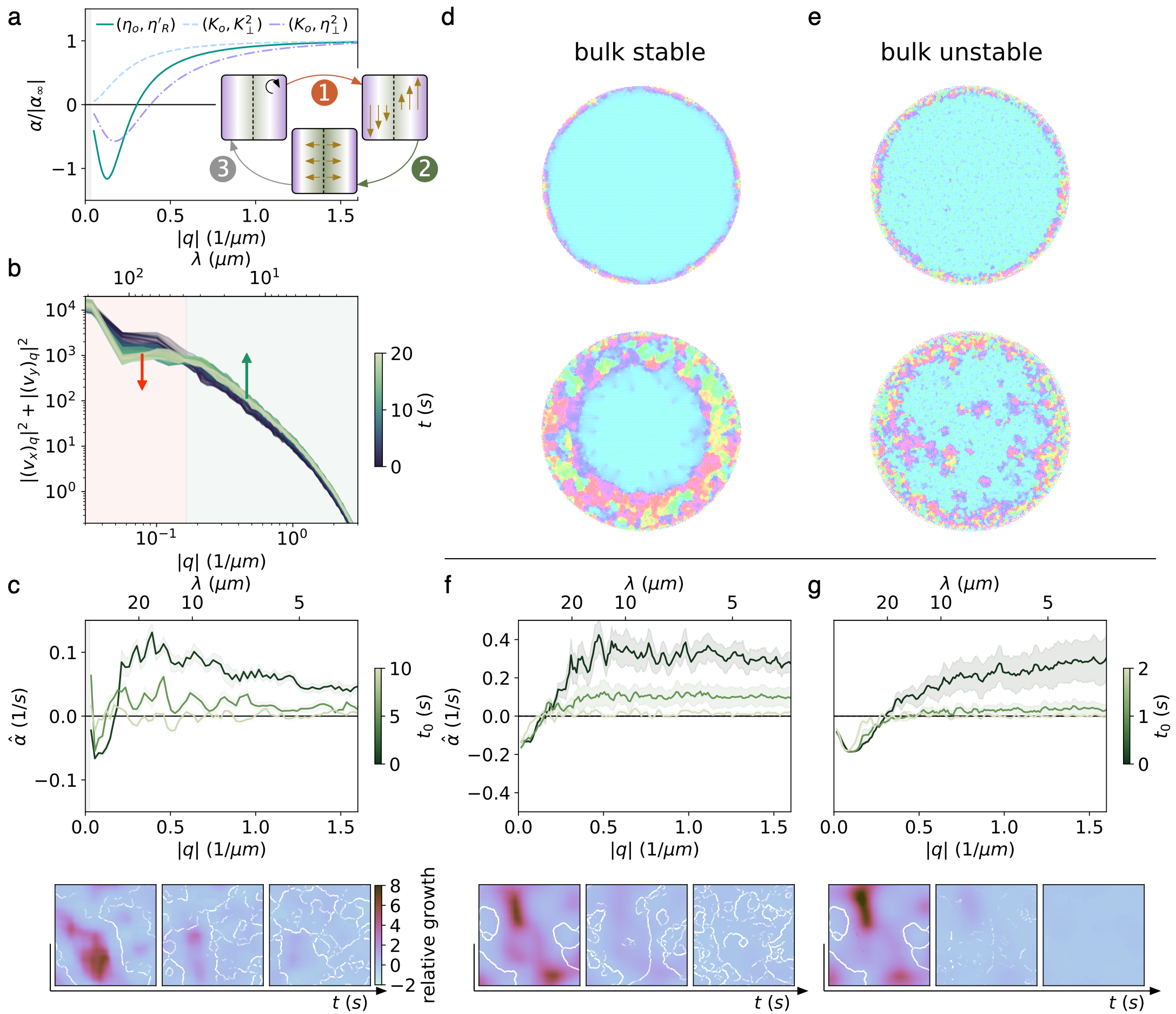}
\caption{\label{fig:instability} {\bf Measuring an elasto-hydrodynamic instability.}
{\bf a,} The chiral phase responds to a perturbation in spinner density by (1) coupling internal rotations to flow gradients that are (2) transformed into Hall stresses that (3) destabilize the crystal further (SI Sec. 8.1). A linear stability analysis of this model yields the predicted spectral growth curves arising from the interplay between odd transport coefficients $(\eta_o, K_o)$ and Magnus-like couplings $(\eta'_R, K_\perp^2, \eta_\perp^2)$.
{\bf b,} As the chiral phase approaches the steady-state of Figure~\ref{fig:polycrystal}, we can measure the spectrum of its flow to see decay at large scales and growth at small scales.
{\bf c,} The total amount of growth measured at these smaller scales can be mapped over time and compared to the grain boundaries to reveal enhanced growth within initially crystalline patches and vanishing growth at later times (SI Sec. 2.4). The estimated spectrum of growth $\hat{\alpha}(q)$ reveals stabilization at scales $\lambda \gg \braket{\xi_\theta}$ and constant destabilization at scales $\lambda \ll \braket{\xi_\theta}$.
{\bf d,} In simulations, we tune the inter-particle interaction range to prepare a theoretically stable crystal, for which the crystal is destabilized through dislocation production to reach the whorl state.
{\bf e,} Likewise, simulations that are theoretically unstable can go unstable in response to bulk density fluctuations. Same colormap as in Figure~\ref{fig:polycrystal}b.
{\bf f,} The estimated spectrum and spatial map of growth for a theoretically stable simulation resembles the experiment when a polycrystal is destabilized. 
{\bf g,} The same signatures are observed in a linearly unstable simulation. {\bf f-g} use the same colormap as {\bf c}.}
\end{figure*}

Linearizing about a homogeneous quiescent base state $\bm{u}=\bm{v}=0$ and $\rho = \rho_0$ and making the ansatz  $\bm{u},\bm{v},\rho \propto \exp{\left(-i\omega t + i\bm{k}\cdot\bm{r}\right)}$, we readily obtain an expression for the dispersion $\nu \equiv \mathrm{Re}(\omega)$ and damping $\alpha\equiv \mathrm{Im}(\omega)$ of displacement and density waves (SI Sec.~8.2). 
We find a generic scenario yielding exponential amplification of density fluctuations.
A number of different combinations of off-diagonal material parameters in $\bm \eta$ and $\bm K$ result in different instabilities. However, they all reflect the same mechanistic picture, sketched in Figure~\ref{fig:instability}a.
A density fluctuation is converted to a localized rotation. 
The resulting net shear across the density fluctuation is in turn converted into an outward force amplifying the initial perturbation and so on to defect unbinding.
Competition with the stabilizing influence of the conventional bulk and shear moduli of the elastic solid determine the consistent shape of the dispersion curves shown in Figure~\ref{fig:instability}a and SI Sec.~8.2.
Crucially, this generic mechanism relies on the coupling between stresses and strains having different spatial symmetries, which is only allowed when time reversal and parity symmetries are broken at the microscopic level~\cite{scheibner2020odd,Epstein2020}.  

To test this simplified model, we measure the flows $\bm{v}(\bm{x})$ that occur in the bulk of a large crystallite  driven at finite rotation frequency (see Supplementary Movie 10), and measure the Fourier spectrum of all scalar measures of deformation, including velocity and strain rate (SI Sec.~2.4). 
Figure~\ref{fig:instability}b shows that following the onset of rotations the azimuthally averaged spectra evolve before eventually settling into a steady state. 
By comparing the spectra at different times, we extract the mode growth curves shown in Figure~\ref{fig:instability} for (c) experiments and (f, g) simulations (SI Sec~2.4, 4.8, 5.4). 
They reveal the presence of an instability at finite wavelength for all modes with relatively constant growth above a characteristic scale.  
As shown in Figure~\ref{fig:instability}c, a spatial map of the integrated growth rate reveals that this instability is consistent with the destabilization of our chiral crystal.

We note, however, that in finite crystalline domains, defects are readily produced at the boundary. 
Their subsequent  propulsion into the bulk could curtail or otherwise affect the  full development of the instability. 
Crucially, we have not observed the spontaneous appearance of defects in the middle of a crystallite in our experiments. 
These combined observations suggest the existence of an additional  mechanism for destabilization in which motile defects nucleate at the boundary and invade the crystal phase, actively fracturing the bulk into whorls.

Our minimal model simulations provide an ideal arena to investigate this possibility. 
By varying microscopic interaction parameters, we can tune the system from linearly stable to linearly unstable (SI Sec.~4.8.1),
Remarkably, in both the  stable and the unstable regime, we observe that initially perfectly crystalline droplets are destabilized by the production of defects at the edges of the droplets, which subsequently invade the bulk in a visible front (Figure~\ref{fig:instability}d-e).
The only difference is that deep in the unstable regime (e), we also observe defect nucleation in the bulk of the droplet before the front arrives. 
Notably, in both cases a crystal whorl emerges as the steady state (SI Sec.~4.8.2). 
This mechanism of destabilization has a different origin from the linear instability described above;  within our continuum approach the presence of a constant background odd stress plays no role in the crystal's stability, rather its primary effect is to  drive defect propulsion (SI Sec.~7.3, 8.2.1).

Remarkably, measurements of the spectral growth in both regimes, performed for simulations initialized in a polycrystalline phase, display general shapes and spatial maps that are similar to those observed in the experiment (Figure~\ref{fig:instability}f-g).
The similar shape of the resulting curves demonstrates the challenge of disentangling the precise origin of the growth in terms of linear response coefficients. 

Breaking parity by spinning a material's constituents gives rise to transverse forces that fundamentally alter the organization of matter. 
Spinner crystals generically melt into a dynamical state driven by  motile dislocations. 
The resulting, tunable crystal-whorl state opens new avenues for control of structure and transport from synthetic materials to biological colonies. 



\noindent
{\bf Acknowledgments.} 
We would like to acknowledge discussions with P. Wiegmann, A. Abanov, D. Nelson, C. Scheibner, M. Han,  M. Fruchart, S. Gokhale, N. Fakhri and J. Dunkel. We thank V. Vitelli for an insightful discussion on the importance of odd stress on defect motility. We thank Wen Yan for useful conversations.

This work was primarily supported by the University of Chicago Materials Research Science and Engineering Center, which is funded by National Science Foundation under award number DMR-2011854.
Additional support was provided by NSF DMR-1905974, NSF EFRI NewLAW 1741685 and the Packard Foundation. 
M.J.S. acknowledges the support from NSF grants  DMR-1420073 (NYU-MRSEC) and DMR-2004469. 
D. B. acknowledges the support from ARN grant WTF and IdexLyon Tore. E.S.B. was supported by the National  Science  Foundation  Graduate  Research  Fellowship  under  Grant  No. 1746045. D.B. and W.T.M.I. gratefully acknowledge the Chicago-France FACCTS programme.  F.B.U. acknowledges support from ``la Caixa'' Foundation (ID 100010434), fellowship LCF/BQ/PI20/11760014, and from the European Union's Horizon 2020 research and innovation programme under the Marie Skłodowska-Curie grant agreement No 847648. 
The  University of Chicago’s Research Computing Center and the University of Chicago's GPU-based high-performance computing system (NSF DMR-1828629) are gratefully acknowledged for access to computational resources and the  Chicago MRSEC (US NSF grant  DMR-2011854) is also gratefully acknowledged for access to its shared experimental facilities.

{\bf Note to be added in proof.} In the concluding stages of our work we became aware of a complementary, independent  effort by the groups of N. Fakhri and J. Dunkel who studied parity breaking crystal dynamics in a bio-physical system (unpublished).

\bibliographystyle{naturemag_nourl}

\end{document}